\documentclass[twoside]{article}
\usepackage{fleqn,espcrc2}

\usepackage{graphicx}
\usepackage[figuresright]{rotating}

\newcommand{\AmS}{{\protect\the\textfont2
  A\kern-.1667em\lower.5ex\hbox{M}\kern-.125emS}}

\title{Heavy-quark condensate at zero and finite temperatures for various forms of the 
short-distance potential}

\author{D. Antonov\thanks{Permanent address: ITEP, B. Cheremushkinskaya 25, 
RU - 117 218 Moscow, Russia.}\address{Institute of Physics,\\ 
 Humboldt University of Berlin, \\
 Newtonstr. 15, D-12489 Berlin, Germany}}

\begin{document}

\begin{abstract}
With the use of the world-line formalism,
the heavy-quark condensate in the SU(N)-QCD is evaluated for the cases when the 
next-to-$1/r$ term in the quark-antiquark potential at short distances is either 
quadratic, or linear. In the former case, which takes place in the stochastic
vacuum model, the standard QCD-sum-rules result 
is reproduced. In the other case, the condensate turns out to be  
UV-finite only in less than four
dimensions. This fact excludes a possibility to have, in four dimensions, the linear term in the 
potential, as well as short strings, at the distances  
smaller than the vacuum correlation length. 
The use of the world-line formalism enables one to generalize further both results for the condensate to the 
case of finite temperatures. A generalization of the QCD-sum-rules result to the case of 
an arbitrary number of space-time dimensions
is also obtained and turns out to be UV-finite, provided this number is smaller than six.
\vspace{1pc}
\end{abstract}

\maketitle

\section{INTRODUCTION}
In this talk, we will briefly review the study performed in Ref.~\cite{0}. Its 
goal was to evaluate, at 
zero and finite temperatures, the heavy-quark condensate in various confining
theories, where the form of the short-distance quark-antiquark potential was
different. 
This can be done by virtue of the formula

$$\left< \bar \psi \psi \right>=-\frac{1}{V}
\frac{\partial}{\partial m} \left< \Gamma\left[A_\mu^a\right]
\right>,$$
where $m$ is the current quark mass, $V$ is the four-volume occupied by the
system, $\left<\ldots\right>$ is defined with respect to the Euclidean Yang-Mills action,
and the one-loop quark self-energy, i.e. the averaged one-loop 
effective action of a spin-$\frac12$ quark reads~\cite{1} (see~\cite{2} for a
recent review):

$$
\left<\Gamma\left[A_\mu^a\right]\right>=
-2\int\limits_{\Lambda^{-2}}^{\infty}\frac{dT}{T}{\rm e}^{-m^2T}
\int\limits_{P}^{} {\cal D}x_\mu
\int\limits_{A}^{} {\cal D}\psi_\mu\times$$

$$\times
\exp\left[-\int\limits_{0}^{T}d\tau\left(\frac14\dot x_\mu^2+
\frac12\psi_\mu\dot\psi_\mu\right)\right]\Biggl\{\Biggl<{\rm tr}{\,}{\cal P}\exp\Biggl[$$

\begin{equation}
\label{2}
ig\int\limits_{0}^{T}d\tau\left(A_\mu\dot x_\mu-\psi_\mu\psi_\nu
F_{\mu\nu}\right)\Biggr]\Biggr>-N\Biggr\}.
\end{equation}
Here, $\Lambda$ stands for an UV momentum cutoff, the subscripts $P$ and $A$ imply the periodic and antiperiodic 
boundary conditions respectively,
$\psi_\mu$'s are antiperiodic Grassmann functions (superpartners of $x_\mu$'s), and 
$A_\mu\equiv A_\mu^a T^a$ with $T^a$'s standing for the 
generators of the SU(N)-group in the fundamental 
representation.  

To calculate the path integral~(\ref{2}), we will transform it to the one in an effective {\it constant} Abelian field, to be 
averaged over. The weight of the average is prescribed by the form of the heavy-quark 
fundamental Wilson loop,
$\left<W(C)\right>\equiv
\left<{\rm tr}{\,}{\cal P}\exp\left(ig\int\limits_{0}^{T}
d\tau A_\mu\dot x_\mu\right)\right>$,
in the {\it original} confining theory.
For heavy (namely, $c$, $b$, and $t$) quarks, $\sqrt{S_{\rm min}(C)}$ is
smaller than the  
vacuum correlation length, where $S_{\rm min}(C)$ is the area of the minimal
surface bounded by the contour $C$.
For such small-sized Wilson loops, one can write 
$S_{\rm min}^2\simeq\frac12\Sigma_{\mu\nu}^2$. Here, $\Sigma_{\mu\nu}=\oint\limits_{C}^{}x_\mu dx_\nu$ is
the so-called tensor area, and 
``$\simeq$'' means ``for nearly flat contours'', that is a reasonable
approximation for a heavy quark, as will be justified below. 
Then, in various confining theories, the following formulae for a small-sized
Wilson loop exist: 

$\bullet$ Stochastic vacuum model (SVM)~\cite{6}, \cite{rev}:

\begin{equation}
\label{3}
\left<W(C)\right>\simeq N\exp\left(-\gamma\Sigma_{\mu\nu}^2\right),
\end{equation}
where $\gamma\equiv\frac{g^2}{8N(d^2-d)}
\left<F^2\right>$.

$\bullet$ Theories with confinement of the Abelian type:

\begin{equation}
\label{4}
\left<W(C)\right>\simeq N\exp\left(-\sigma\sqrt{\frac12
\Sigma_{\mu\nu}^2}\right)
\end{equation}
with the following values of $\sigma$:

-- in the SU(N)-version of the weakly coupled 3d Georgi-Glashow model~\cite{7}
$\sigma=\frac{\pi}{2}\frac{N-1}{\sqrt{N}}g\sqrt{\zeta}$. Here,
$\zeta\sim\frac{M_W^{7/2}}{g}{\rm e}^{{-4\pi\varepsilon M_W}/g^2}$
is the
monopole fugacity, $1\le\varepsilon < 2$, $g$ is the electric coupling
constant, and $M_W$ is the W-boson mass, $g^2\ll M_W$;

-- in the London limit of the 4d SU(N)-inspired dual Abelian-Higgs--type theory~\cite{8}
$\sigma=2\pi(N-1)\eta^2\ln\kappa$, where $\eta$ is the vacuum expectation
value of the dual Higgs field, and $\kappa$ is the Landau-Ginzburg parameter,
$\ln\kappa\gg 1$.

Therefore, under the assumption that the Feynman-Kac formula, which relates the
potential of a heavy quark-antiquark pair to the Wilson loop, can be
extrapolated up to the distances of the order of the vacuum correlation
length, the next-to-$1/r$ term of the potential is $\propto r^2$ in the
SVM, whereas it is $\propto r$ in the Abelian-type
theories with confinement.

\section{HEAVY-QUARK CONDENSATE AT ZERO TEMPERATURE}
In the SVM at arbitrary $d\ge 2$, one has by virtue of eqs.~(\ref{2}),
(\ref{3}):

$$\left<\Gamma\left[A_\mu^a\right]\right>=
-\frac{2NV}{(8\pi\gamma)^{n/2}}\int\limits_{\Lambda^{-2}}^{\infty}\frac{dT}{T}{\rm
  e}^{-m^2T}\times$$

$$\times
\left(\prod\limits_{\mu<\nu}^{}\int\limits_{-\infty}^{+\infty}dB_{\mu\nu}
{\rm e}^{-\frac{B_{\mu\nu}^2}{8\gamma}}\right)\left\{
\int\limits_{P}^{} {\cal D}x_\mu
\int\limits_{A}^{} {\cal D}\psi_\mu\times\right.$$

$$\times\exp\left[-\int\limits_{0}^{T}d\tau\left(\frac14\dot x_\mu^2+
\frac12\psi_\mu\dot\psi_\mu+\frac{i}{2}B_{\mu\nu}x_\mu\dot x_\nu-\right.\right.$$

\begin{equation}
\label{5}
-iB_{\mu\nu}\psi_\mu\psi_\nu\Bigr)\Biggr]-\frac{1}{(4\pi T)^{d/2}}
\Biggr\},
\end{equation}
where $n=\frac{d^2-d}{2}$ is one half of the number of off-diagonal components
of the space-time independent field $B_{\mu\nu}$.
The expression in the curly brackets can further be recognized as the 
Euler-Heisenberg-Schwinger Lagrangian, whose small-$T$ (large-$m$) expansion
yields $\left\{\ldots\right\}=\frac{1}{(4\pi T)^{d/2}}\left[
\frac{T^2}{3}\sum\limits_{\alpha<\beta}^{}B_{\alpha\beta}^2+{\cal O}
\left(T^4\left(B_{\mu\nu}^2\right)^2\right)\right]$.
The actual parameter of this expansion is
$\frac{g^2\left<F^2\right>}{N\left(d^2-d\right)m^4}$ (it is ${\cal O}(N^0)$), 
and the neglected term is 
$\sim\left(\frac{g^2\left<F^2\right>}{N\left(d^2-d\right)m^4}\right)^2$. 
For $d\sim 4$, the expansion holds for $c$, $b$, and $t$ quarks, but not for 
$u$, $d$, and $s$ quarks.

One can further see that
$\left<\bar\psi\psi\right>$ diverges as $\ln\frac{\Lambda}{m}$ for $d=6$ and as 
$(\Lambda/m)^{d-6}$ for $d>6$, while for $d<6$ it is finite and reads

$$
\left<\bar\psi\psi\right>=-\frac{m^{d-5}\Gamma\left(3-\frac{d}{2}\right)}{3(4\pi)^{\frac{d}{2}-1}}
\alpha_s\left<F^2\right>.
$$
In particular,
$\left.\left<\bar\psi\psi\right>\right|_{d=4}=-\frac{\alpha_s
\left<F^2\right>}{12\pi m}$,
that coincides with the result of Ref.~\cite{shif},
justifying the approximation 
$S_{\rm min}^2\simeq\frac12\Sigma_{\mu\nu}^2$ for a heavy quark. At $d=2$ and
$d=3$, the results read

$$
\left.\left.\left<\bar\psi\psi\right>\right|_{d=3}=\frac{m}{4}\left<\bar\psi\psi\right>\right|_{d=2}=
-\frac{\alpha_s\left<F^2\right>}{12m^2}.
$$

In the theories with the Abelian-type confinement, at arbitrary $d\ge 2$, we
have

$$\left<\Gamma\left[A_\mu^a\right]\right>=
-\frac{2NV\Gamma\left(\frac{n+1}{2}\right)}{\pi^{\frac{n+1}{2}}\sigma^n}
\int\limits_{\Lambda^{-2}}^{\infty}\frac{dT}{T}{\rm e}^{-m^2T}\times$$

$$\times
\left(\prod\limits_{\mu<\nu}^{}\int\limits_{-\infty}^{+\infty}dB_{\mu\nu}\right)
\frac{1}{\left(1+\frac{1}{2\sigma^2}B_{\mu\nu}^2\right)^{\frac{n+1}{2}}}
\left\{\ldots\right\},$$
that is similar to Eq.~(\ref{5}), but with a different measure of average over $B_{\mu\nu}$.
The expansion of $\left\{\ldots\right\}$ now goes in powers of
$\sigma/m^2$. For $\sigma$ of the order of $(440{\,}{\rm MeV})^2$,
the expansion again holds for $c$, $b$, and $t$ quarks.
One can further see that $\left<\bar\psi\psi\right>$ diverges as 
$(\Lambda/m)^{d-4}$ for $d>4$, while
$\left.\left<\bar\psi\psi\right>\right|_{d=4}=-\frac{5Nm\sigma}{(4\pi)^2}\ln\frac{\Lambda}{m}$.
This divergency contradicts the apparent finiteness of the (nonperturbative
part of the) heavy-quark condensate and necessitates to attribute some
physical meaning to $\Lambda$. 
In the London limit of the 4d SU(N)-inspired dual Abelian-Higgs--type theory, 
$\Lambda$ is of the order of the mass of the dual Higgs boson. 
In 4d QCD, we come to the conclusion that $\left<W(C)\right>$ may not have the form 
${\rm e}^{-\sigma S_{\rm min}(C)}$ up to arbitrarily short distances. There,
$\Lambda$ is of the order of the inverse thickness
of a ``short string''. In terms of the heavy-quark potential, this result
means that the linear next-to-$1/r$ term may exist only up to distances not
smaller than the vacuum correlation length. Very short strings, of a length
much smaller than the vacuum correlation length, are therefore ruled out in 4d QCD, as
well as the linear term in the potential at such distances.

Instead, at $d<4$, $\left<\bar\psi\psi\right>$ is finite:

$$\left<\bar\psi\psi\right>=-\frac{N\Gamma\left(\frac{n+1}{2}\right)\Gamma\left(2-\frac{d}{2}\right)
\sigma m^{d-3}}{3\cdot 2^{d-3}\Gamma\left(\frac{n}{2}\right)
\pi^{\frac{d+1}{2}}}.
$$
In particular,
$\left.\left.\left<\bar\psi\psi\right>\right|_{d=3}=m\left<\bar\psi\psi\right>\right|_{d=2}
=-\frac{2N\sigma}{3\pi^2}$.

\section{FINITE-TEMPERATURE GENERALIZATIONS}
In this section, we will consider the finite-temperature generalizations of
the above-obtained results, starting with the SVM at $d=4$.
Antiperiodic boundary conditions for quarks can be taken into account upon the multiplication 
of the zero-temperature heat kernel by the factor
$\left[1+2\sum\limits_{n=1}^{\infty}(-1)^n{\rm e}^{-\frac{\beta^2n^2}{4T}}\right]$, where $\beta=1/({\rm temperature})$~\cite{bras}.
One has

$$\left<\bar\psi\psi\right>=
\left<\bar\psi\psi\right>_0
\left[1+2\beta m\sum\limits_{n=1}^{\infty}(-1)^nnK_1(\beta mn)\right],
$$
where $\left<\bar\psi\psi\right>_0=-\frac{\alpha_s
\left<F^2\right>}{12\pi m}$, and
$\left<F^2\right>$ experiences a drop by a factor of the order of 2 at the
deconfinement temperature, $T_c\sim 150{\,}{\rm MeV}$,
due to the evaporation of the chromoelectric condensate~\cite{evap}~\footnote{
Henceforth, $T$ will denote the temperature, rather than the proper time.}.
For $c$, $b$, and $t$ quarks, $T_c\ll m$, and  
only the first Matsubara mode yields a sufficient contribution:
$\left<\bar\psi\psi\right>\simeq\left<\bar\psi\psi\right>_0
\left(1-\sqrt{2\pi\beta m}{\rm e}^{-\beta m}\right)$. The correction to the
zero-temperature result is therefore exponentially small. Next, 
when $T$ starts exceeding the temperature of dimensional reduction, $T_{\rm d.r.}$,
the theory becomes three-dimensional.
Since, $T_{\rm d.r.}$ is not larger than $2T_c$ (see e.g.~\cite{rev}), 
for $c$, $b$, and $t$ quarks a rather broad range of temperatures above $T_{\rm d.r.}$ exists, where $T\ll m$.
At such temperatures, one gets
$\left<\bar\psi\psi\right>\simeq-\frac{\alpha_s\left<F^2\right>T}{12m^2}
\left(1-2\beta m{\rm e}^{-\beta m}\right)$.

In the Abelian-type theories with confinement, one should first of all note
that the radius of a ``short string'', $r_\perp$, grows with the
temperature. Therefore, one may substitute  $\Lambda$ by $ r_\perp^{-1}$ only 
as long as $mr_\perp(T)\ll 1$. For the sake of generality, let us assume that this is true up to 
temperatures larger than $T_{\rm d.r.}$, and that $T_{\rm d.r.}>T_c$ as in
QCD. Implying everywhere below the above-mentioned substitution $\Lambda\to r_\perp^{-1}$, we have at $T<T_c$,

\begin{equation}
\label{conden}
\left<\bar\psi\psi\right>\simeq\left<\bar\psi\psi\right>_0
\left(1-\sqrt{\frac{2\pi}{\beta m}}\frac{{\rm e}^{-\beta
      m}}{\ln\frac{\Lambda}{m}}\right),
\end{equation} 
where $\left<\bar\psi\psi\right>_0=-\frac{5Nm\sigma}{(4\pi)^2}\ln\frac{\Lambda}{m}$.
At $T>T_c$, only the spatial string tension, $\sigma_s$, does not
vanish, and the Wilson loop takes the form 
$\left<W(C)\right>\simeq N\exp\left[-\sigma_s\left(\Sigma_{12}^2+\Sigma_{13}^2+\Sigma_{23}^2\right)^{1/2}\right]$.
Then, at $T\in\left[T_c, T_{\rm d.r.}\right]$, $B_{\mu\nu}$ is a $3\times 3$-matrix, but 
the theory is still four-dimensional. These two facts together yield the
following formula for the condensate:

\begin{equation}
\label{mixed}
\left<\bar\psi\psi\right>\simeq
-\frac{2Nm\sigma_s}{3\pi^3}\ln\frac{\Lambda}{m}\cdot
\left(1-\sqrt{\frac{2\pi}{\beta m}}\frac{{\rm e}^{-\beta m}}{\ln\frac{\Lambda}{m}}\right).
\end{equation}
The difference of the overall factor $\frac{2Nm\sigma_s}{3\pi^3}$ from the factor
$\frac{5Nm\sigma}{(4\pi)^2}$ of Eq.~(\ref{conden}) is due to the fact that the temporal string tension
vanishes when one passes from $T<T_c$ to $T>T_c$. Naively, one could have expected at $T=T_c+0$ the factor 
$\frac{5Nm\sigma_s}{(4\pi)^2}$, that is not the case. Finally, at 
$T>T_{\rm d.r.}$, the theory becomes fully three-dimensional. There, 
$\left<\bar\psi\psi\right>\simeq-\frac{2N\sigma_sT}{3\pi^2}
\left(1-2{\rm e}^{-\beta m}\right)$, as 
long as $T\ll m$.

\section{SUMMARY}
By making use of the world-line formalism, we have evaluated the heavy-quark
condensate at zero and finite temperatures.
The respective heavy-quark Wilson loop has been considered either within the SVM, or 
within the theories with Abelian-type confinement. The zero-temperature
results are the following:

$\bullet$ SVM, $d=4$: the result of Ref.~\cite{shif} is reproduced; $\left<\bar\psi\psi\right>$
diverges as $\ln\frac{\Lambda}{m}$ at $d=6$ and as 
$(\Lambda/m)^{d-6}$ at $d>6$, while the finite expression is obtained at $2\le d <6$.

$\bullet$ Theories with Abelian-type confinement:
the finite expression is obtained at $2\le d <4$, while 
$\left<\bar\psi\psi\right>$ diverges as 
$(\Lambda/m)^{d-4}$ at $d>4$ and as $\ln\frac{\Lambda}{m}$ at $d=4$.
Possible physical meanings of the UV cutoff at $d=4$:

-- 4d SU(N)-inspired dual Abelian-Higgs--type theory: $\Lambda$ is of the order of
the mass of the dual Higgs boson; 

-- If, in some approximation, QCD can also be considered as a theory belonging
to this class, then $\Lambda$ is the inverse thickness
of a ``short string''. Therefore, the
``short string'' and the associated 
linear next-to-$1/r$ term in the heavy-quark
potential may exist only up to distances not smaller than the vacuum correlation length.

Finally, the finite-temperature generalizations of the above-discussed results have
also been presented. In general, antiperiodic boundary conditions for quarks produce corrections,
which are exponentially small in the parameter $\beta m$ (both below the temperature of dimensional
reduction and in the broad range of temperatures above it). A nontrivial situation appears, in case of the 
short-distance linear potential, at temperatures lying between the deconfinement critical temperature and the 
temperature of dimensional reduction. Although the theory is still four-dimensional in this phase, the 
Wilson loop is that of a 3d theory (since only the spatial string tension survives the 
deconfinement phase transition). The result for the condensate in this phase is given by Eq.~(\ref{mixed}).

\section*{ACKNOWLEDGMENTS}
I am grateful to the Alexander von Humboldt foundation for the financial
support and to the organizers of the conference "QCD 04" (Montpellier, France, 5-9 July, 2004)
for an opportunity to present these results in a very stimulating
atmosphere.

\end{document}